# Predicting the Oxidation States of Mn ions in the Oxygen Evolving Complex of Photosystem II Using Supervised and Unsupervised Machine Learning

Muhamed Amin*[1,2,3]

[1] Department of Sciences, University College Groningen, University of Groningen, Hoendiepskade 23/24, 9718 BG Groningen, The Netherlands.

[2] Rijksuniversiteit Groningen Biomolecular Sciences and Biotechnology Institute, University of Groningen, Groningen, Netherlands

[3] Center for Free-Electron Laser Science, Deutsches Elektronen-Synchrotron DESY, Notkestrasse 85, 22607 Hamburg, Germany.

m.a.a.amin@rug.nl

**Abstract**

Serial Femtosecond Crystallography at the X-ray Free Electron Laser (XFEL) sources enabled the imaging of the catalytic intermediates of the oxygen evolution reaction of Photosystem II. However, due to the incoherent transition of the S-states, the resolved structures are a convolution from different catalytic states. Here, we train Decision Tree Classifier and K-mean clustering models on Mn compounds obtained from the Cambridge Crystallographic Database to predict the S-state of the X-ray, XFEL, and CryoEm structures by predicting the Mn's oxidation states in the oxygen evolving complex (OEC). The model agrees mostly with the XFEL structures in the dark $S_1$ state. However, significant discrepancies are observed for the excited XFEL states ($S_2$, $S_3$, and $S_0$) and the dark states of the X-ray and CryoEm structures. Furthermore, there is a mismatch between the predicted S-states within the two monomers of the same dimer, mainly in the excited states. The model suggests that improving the resolution is crucial to precisely resolve the geometry of the illuminated S-states to overcome the noncoherent S-state transition. In addition, significant radiation damage is observed in X-ray and CryoEM structures, particularly at the dangler Mn center (Mn4). Our model represents a valuable tool for investigating the electronic structure of the catalytic metal cluster of PSII to understand the water splitting mechanism.

**Introduction**

Throughout history, Nature has inspired humans and driven many discoveries and inventions. For scientists, Nature is a vast school in which to observe, record, learn, and get inspired. Likewise, remarkable technologies and products have been inspired in one way or another by lessons learned from the environment, from sailing to flying to Velcro. Similarly, understanding the machinery of the biological nano-engines responsible for Photosynthesis would help understand how common metals such as manganese (Mn) act within Photosystem II (PSII) as the best catalyst for water oxidation (Fig. 1).[1] A comprehensive understanding of the machinery can pave the way for developing similar artificial catalysts[2-6]. After recognizing the need to transition to renewable energy sources, Giacomo Ciamician proposed that photochemical devices can convert solar energy into fuel.[7] This idea was one of the early motivations to develop artificial Photosynthesis.

In natural Photosynthesis, PSII, a membrane protein complex in cyanobacteria, algae, and higher plants, harvests solar energy to drive water oxidization, converting the light energy into chemical energy and releasing di-molecular oxygen as a byproduct. In the '70s of the 20$^{th}$ century, Bassel Kok described the biological water oxidation process in a five-step reaction[8]. PSII carries out this reaction by coupling four-electron water oxidation at the oxygen-evolving complex (OEC), with the one-electron photochemistry occurring at the reaction center[9-11]. The OEC consists of a heteronuclear $Mn_4O_5Ca$ cluster, and it cycles through five intermediate S-states ($S_0$ to $S_4$) that corresponds to the abstraction of four successive electrons from the OEC[8]. Several cofactors, i.e., chlorophyll, pheophytin, quinones, non-heam iron, and a redox-active tyrosine sidechain, are involved in the charge separation reaction during the water oxidation (Fig. 1).

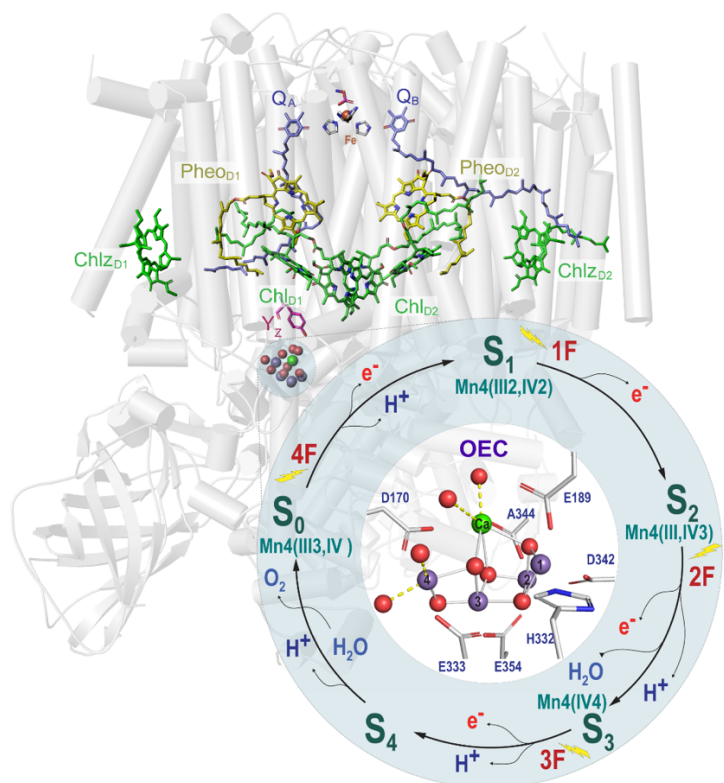

**Fig. 1** The structure of the monomeric PSII with all the subunits is shown as a cartoon in gray. All the redox-active cofactors are involved in charge transfer are shown; the manganese ions of the OEC depicted in purple, the chlorophyll (Chl) in green, the pheophytin (Pheo) in yellow, the non-heme iron (Fe) in red, the quinones (Q) in light blue, and the tyrosine ($Y_z$) in magenta. On the bottom right, Kok's cycle of the oxygen evolution that takes place at the OEC of the PSII is shown. It shows the steps of the water oxidation reaction that is triggered by the absorption of photons shown as five oxidation states ($S_0 \rightarrow S_4$)

A complete understanding of the catalytic activity of the PSII requires detailed information about the geometric and the electronic structure of the $Mn_4CaO_5$ cluster. The atomic geometric structure was first revealed at a resolution of 1.9Å in 2011, using synchrotron X-ray crystallography[12]. However, synchrotron radiation induces sample damage that can influence the geometric structure of the active metal site[13-15]. On the other hand, the recent advancement in the X-ray Free-electron Laser (XFEL) crystallography provided a radiation damage-free geometric structure of the $Mn_4CaO_5$ cluster[16-21]. Moreover, XFEL studies provided the geometric structure for the $S_1$ state (dark-adapted) and other S states at a resolution of ~ 2.0Å. However, despite the advancements in understanding the geometric structure of the $Mn_4CaO_5$ cluster, the electronic structure is still elusive.

In this study, we built machine learning models to predict the oxidation states of Mn in the OEC and hence the S-state of the X-ray, XFEL, and cryoEM structures. The model is trained on Mn-containing small molecules obtained from the Cambridge Crystallographic Database (CCD), where the oxidation states are already known. The models, which showed very high accuracy scores (above 95%) on the training dataset, agreed mostly with the XFEL structures in the dark-adapted state ($S_1$). However, significant discrepancies are observed for the X-ray and cryoEM $S_1$ structures, as well as the illuminated XFEL structures ($S_2$, $S_3$, and $S_0$).

**Results and Discussion**

To predict the oxidation states of the Mn in the oxygen evolution complex (OEC) of Photosystem II, we built a prediction model based on the data we collected for Mn compounds from the Cambridge Crystallography Database, where the oxidation states of the Mn are already known. Only small compounds with crystallographic data, R-factors ≤ 0.075, and are error-free (at the level of 0.05Å) were included in the search[22]. Furthermore, only octahedral Mn compounds with oxygen (O) and nitrogen (N) ligands were selected because the Mn ions in the OEC are coordinated by O and N ligands. In total, the database that was built contains 1734, 835, and 107 structures corresponding to the oxidation states Mn(II), Mn(III), and Mn(IV), respectively.

The average bond lengths between the Mn and the ligands are significantly different for the different oxidation states; shorter for higher Mn oxidation states. Furthermore, in the case of Mn(III), the axial ligands have a significantly longer bond length due to the Jahn Teller effect. Therefore, the prediction model was designed to assess two features: 1) The average bond length between the Mn and the equatorial ligands 2) and the axial ligands. Fig. 2a shows the average bond lengths of the equatorial ligands (X-axis) against the axial ligands (Y-. Although there are clear distinguished clusters for Mn(II) (Cyan), Mn(III) (Blue), and Mn(IV) (Dark Blue), there are some mislabeled elements within each cluster. There are several reasons for the mislabeled data, such as inter-ligand steric effects.[23] In addition, the asymmetry between the X and Y axes is observable in the Mn(III) cluster due to the Jahn Teller distortion.

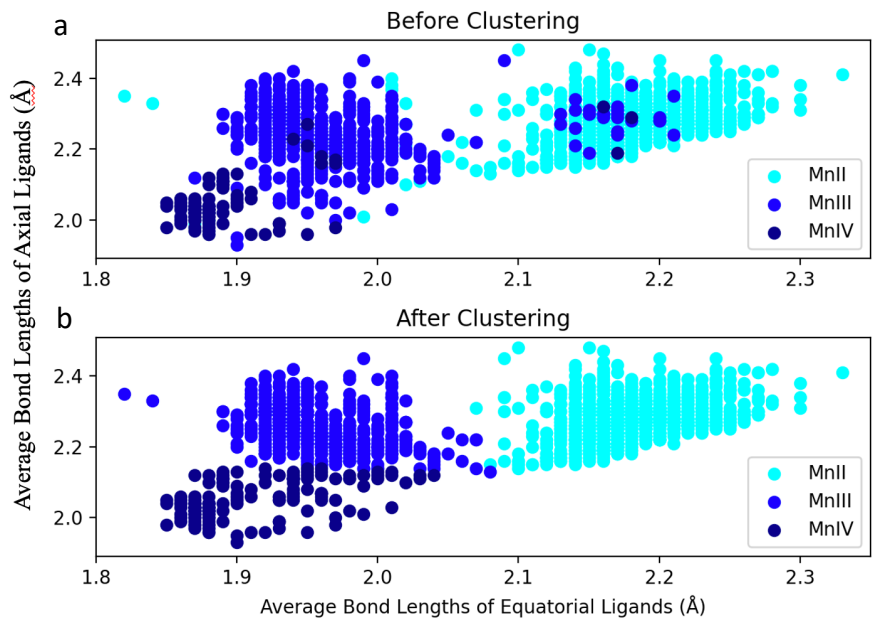

**Fig. 2** The average bond lengths of the equatorial ligands against axial ligands a) Raw data before clustering b) The adjusted data after K-mean clustering.

The *K*-mean clustering algorithm was used to label our data into three distinct clusters to correct the mislabeled data. As shown in Fig. 2b, the algorithm converged, and the data were clustered into three clusters with the following centers: (2.18, 2.28) Å for Mn(II), (1.95, 2.26) Å for Mn(III), and (1.91, 2.05) Å for Mn(IV). The ratio Y/X for the three centers are (1.05, 1.16 1.08) Å, corresponding to Mn(II), Mn(III), and Mn(VI), respectively. The cluster that corresponds to Mn(III) clearly shows the effect of Jahn Teller distortion, where the axial ligands are significantly longer than the equatorial ones. Finally, we used the clustered data to build a prediction model based on two different classifiers: Gaussian Naïve Bayes and Decision Tree classifier.

**Gaussian Naïve Bayes Classifier (GNB)**

The reason for choosing the GNB is that the data could be fitted to 2D Gaussians. Thus, we expected the model to perform well given the training dataset. The model is trained on 75% of the data, and the remaining 25% are used for testing. Before processing the data with K-mean clustering, the accuracy score for the GNB prediction model is 94%, and the confusion matrix that shows the prediction against the true labels is shown in Fig. 3 (upper left). Furthermore, we performed a 10-fold cross-validation to evaluate further the model, which resulted in a mean accuracy score of 96% and a sigma of 1%.

The accuracy score increased to 99% after using the processed data after the clustering, which is also reflected in the confusion matrix shown in Fig. 3 (upper right). Most of the wrong predictions are Mn(IV) data points, which were classified as Mn(III) (6 out of 44). Interestingly, the means of the Gaussians used to calculate the prior probabilities precisely match the center of the clusters obtained from the *K*-mean algorithm.

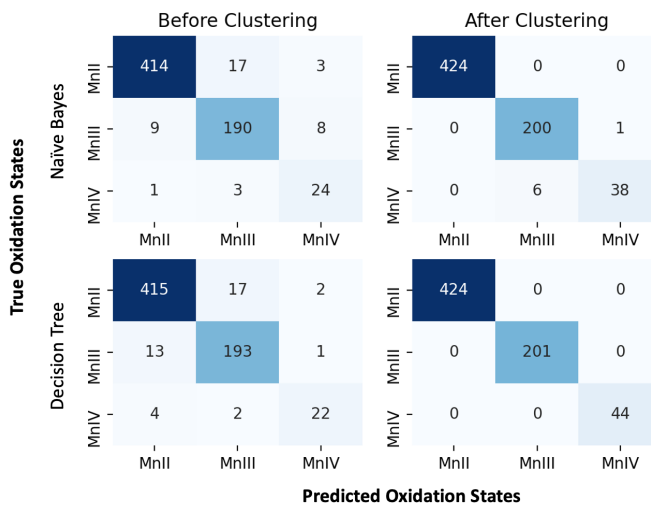

**Fig. 3**. The confusion matrix of GNB (top) and DT (bottom) prediction models before (left) and after (right) k-mean clustering.

**Decision Tree Classifier (DT)**

The Decision Tree (DT) classifier is based on a very different algorithm than the Naïve Bayes. Each node in the tree applies a test on a feature (here, the average bond lengths of the axial or the equatorial ligands); the branches descending from each node correspond to one of the possible values for that feature. The nodes are arranged in the tree so that the reduction in the information entropy is maximized.

The DT model shows a higher accuracy score than the GNB before and after the clustering (Fig. 3 lower left, lower right, respectively). The cross-validation calculations show an accuracy score of 95% before the clustering and 100% after the clustering. The set of rules used to classify the Mn oxidation states are shown in Fig. 4.

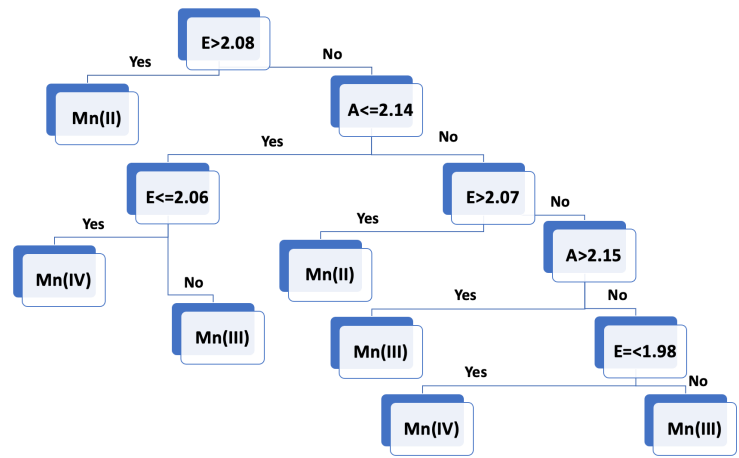

**Fig. 4** The fitted Decision Tree. E and A are the average bond lengths of the equatorial and axial ligands, respectively. The left branches are followed if the condition is satisfied.

**Prediction of the catalytic states of the Kok cycle**

The OEC contains four Mn ions and one Ca ion (Fig. 1). The Mn ions are ligated mainly by (O) and one (N). During the catalytic cycle of the PSII enzyme, the Mn ions are oxidized in the transition between the S-states. According to different spectroscopic studies, using different techniques, the oxidation state of the Mn ions in the dark-adapted state of the OEC ($S_1$-state) is Mn(III, IV, IV, III)[24-27]. The Mn(III) is then oxidized in the transition to higher S-states till all Mn are Mn(IV) in the $S_3$-state (Fig. 1). Our prediction model with the highest accuracy score based on Decision Tree Classifier was used to predict the oxidation states of the Mn in Photosystem II in 38 structures (27 XFEL, 6 X-ray, and 5 cryoEM structures) for each monomer independently. Only the structures of the meta-stables S-states ($S_1$, $S_2$, $S_3$, $S_0$), which have a resolution of 2.5Å or better, were included in this study (Table 1). After the predictions of the oxidation states, the S-states were assigned according to the total charges of the 4-Mn, i.e., the total charges for $S_0$, $S_1$, $S_2$, and $S_3$ are 13, 14, 15, and 16, respectively. In addition, we have included the $S_{-1}$, $S_{-2}$, …, $S_{-5}$ to account for more reduced structures.

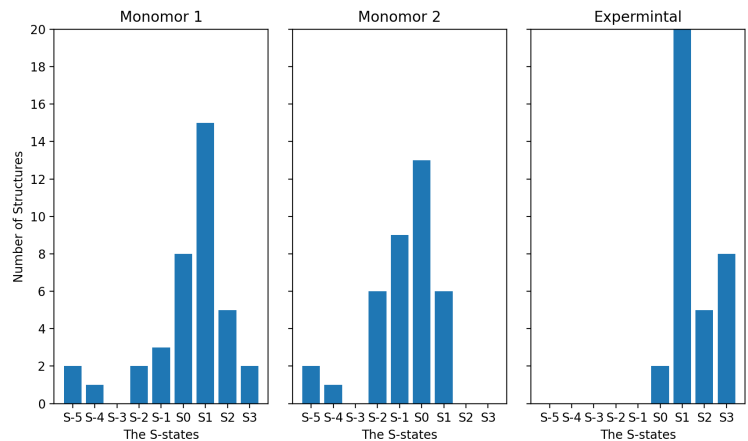

**Fig. 5** A histogram of the predicted (for monomer1 and monomer 2) and the labeled S-states of the 34 structures of the OEC. The S-states are assigned according to the total charges of the 4-Mn in the OEC.

The prediction accuracy of our model is 96% before clustering and nearly 100 % after clustering when applied to the small molecules. To further assess our model, we predicted the oxidation states of the Mn in DFT optimized structures of the OEC, where the oxidation states are already known from the spin densities.[28] The model successfully predicted the oxidation state that matches the Mn's spin densities. On the other hand, the first impression of the prediction model's accuracy to predict the oxidation states of the OEC is low. As shown in Fig. 5, the structures are mostly in the $S_1$ state for both monomers. However, the experimentally assigned S-states, which are based on the number of flashes used to pump the sample, are significantly different from the predicted by our model. For monomer 1, out of the 38 structures, the prediction matched the experimentally assigned S-state for ten structures; all of them are assigned as $S_1$ except for the 6dhf and 6w1v structures, which are $S_2$ and $S_3$, respectively (Table 1). For monomer 2, 10 structures matched the experimentally assigned S-state, all of them for $S_1$ except for the 7rf3 structure, which is $S_2$. (PDB ID: 6dhp).

A closer and more detailed look at the predicted oxidation states shows a totally different story. Among the investigated PSII PDB files, 22 structures correspond to the $S_1$-state (PDB; 3wu2, 4il6, 4pj0, 4ub6, 4ub8, 5b66, 5b5e, 5gth, 5h2f, 5ws5, 5zzn, 6dhe, 6jlj, 6jlm, 6w1o, 7cji, 7cou, 7rf3, 7d1t, 7d1u, 7n8o, 7rcv); eleven out of the 22 have been solved using XFEL data. In seven XFEL-$S_1$ structures, the predicted S-states, using our models, agreed with the experimental one in both monomers. Furthermore, in another three XFEL-$S_1$ structures, the predicted S-stated agreed with the experimental one in at least one of the two monomers. Although, the predicted S-state did not match the reported one in only one XFEL-$S_1$ structure. The agreement between the XFEL structures and the prediction in the dark state ($S_1$) supports that our models accurately predict the damage-free XFEL-$S_1$ structures.

The other 11 structures were solved using data that have been collected using conventional synchrotron X-ray radiation (PDB; 3wu2, 4il6, 4pj0, 5b66, 5b5e, 5h2f) or cryoEM (PDB; 5zzn, 7n8o, 7rcv, 7d1t, 7d1u). While the predicted and reports states are mostly in agreement, for the XFEL $S_1$-structure, the reported S-states for these 11 structures mainly disagreed with the predicted one. Only one monomer of the 11 structures was predicted to be in the $S_1$-state as reported, while the rest are predicted to be in more reduced states $S_0$, $S_{-1}$, …, or $S_{-5}$ (Table 1). Some of these $S_1$-models were investigated theoretically, and it was predicted to suffer from severe radiation damage.[29, 30] It is worth noticing that the single agreement (in the case of the X-ray $S_1$ structures) is coming from a collected dataset using a low dose of synchrotron radiation.[17] These observations emphasize the influence of the radiation damage on the OEC electronic structure and hence the geometrical structure, in agreement with several studies that generally assess the radiation damage in protein crystallography.[13-15, 31-34] Moreover, several studies have discussed the radiation damage in the case of PSII, particularly during X-ray[35, 36], and the cryoEM data collection.[30]

On the other hand, the agreement of the prediction was relatively low for the XFEL structures (PDB: 5gti, 5tis, 5ws6, 6dhf, 6dho, 6dhp, 6jlk, 6jll, 6jln, 6jlo, 6jlp, 6w1p, 6w1v, 7cjj, 7rf3, 7rf8), which are for different excited S-states, $S_2$, $S_3$, and $S_0$. It could be due to the incoherent S-state transition after illumination, leading to the presence of different S-states within the same structure. Overcoming such a problem requires a higher resolution so that the Mn-O distances are better resolved. However, we can not eliminate the possibility that the presence of the Ca ion in the OEC affects the cluster's geometry during the light activation. The Ca ion in the OEC has been intensively investigated; it plays a critical to the substrate insertion during the light activation.[37-39]

Overall, the predicted oxidation state of the Mn ions for the XFEL structures did not mostly show Mn(II) content, except for PDB 7cou, 7cji, 6jlp, where one of the two monomers contained Mn ion that was predicted to be in Mn(II) oxidation state. Unlike the XFEL structures, most of the synchrotron or cryoEM structures contain Mn(II), and all Mn in the 7n8o, 7rcv structures are reduced to Mn(II), likely because of the high radiation dose.[40] In addition, these structures show higher reductions states, i.e., $S_{-1}$, …, or $S_{-5}$, that are physiologically do not exist in an active PSII, indicating that the suffering of radiation damage significantly influences the electronic and geometrical structure of the OEC. The radiation damage influences the OEC geometry, and it is manifested in the prediction results of the $S_1$-structures. These results emphasize the importance of radiation-free data collection to study a functional PSII.

Interestingly, for all the structures that showed a presence of Mn(II) in the OEC, the Mn(II) oxidation state was permanently assigned to the Mn4 (the dangling Mn), indicating the high vulnerability of this Mn ion. The side chains D170 and E333 ligate Mn4; in the vicinity of D170, it is recently suggested that a Mn(II) high-affinity site, where the Mn ions are oxidized in preparation for the OEC formation during the OEC assembly.[41]

| PDBID | S-state Reported | S-state Monomer 1 | Mn Oxidation Monomer 1 Mn1,Mn2,Mn3,Mn4 | S-state Monomer 2 | Mn Oxidation Monomer1 Mn1,Mn2,Mn3,Mn4 | Flashes (time delay between flashes) |
|---|---|---|---|---|---|---|
| 3wu2[42] | $S_1$ | $S_{-2}$ | Mn(III,III,III,II) | $S_{-1}$ | Mn(III,IV,III,II) | X-ray |
| 4il6[39] | $S_1$ | $S_0$ | Mn(III,IV,III,III) | $S_0$ | Mn(III,IV,III,III) | X-ray |
| 4pj0[43] | $S_1$ | $S_{-1}$ | Mn(III,IV,III,II) | $S_{-2}$ | Mn(III,III,III,II) | X-ray |
| 4ub6[17] | $S_1$ | $S_0$ | Mn(III,IV,III,III) | **$S_1$** | Mn(III,IV,IV,III) | XFEL |
| 4ub8[17] | $S_1$ | $S_0$ | Mn(III,IV,III,III) | $S_0$ | Mn(III,IV,III,III) | XFEL |
| 5b5e[44] | $S_1$ | $S_0$ | Mn(III,IV,III,III) | **$S_1$** | Mn(III,IV,IV,III) | X-ray |
| 5b66[44] | $S_1$ | $S_0$ | Mn(III,IV,III,III) | $S_0$ | Mn(III,IV,IV,II) | X-ray |
| 5gth[45] | $S_1$ | **$S_1$** | Mn(III,IV,IV,III) | **$S_1$** | Mn(III,IV,IV,III) | XFEL |
| 5gti[45] | $S_3$ | $S_1$ | Mn(III,IV,IV,III) | $S_1$ | Mn(III,IV,IV,III) | XFEL |
| 5h2f[46] | $S_1$ | $S_{-1}$ | Mn(III,III,III,III) | $S_{-1}$ | Mn(III,IV,III,II) | X-ray |
| 5tis[47] | $S_3$ | $S_0$ | Mn(III,IV,III,III) | $S_0$ | Mn(III,III,IV,III) | XFEL |
| 5ws5[45] | $S_1$ | **$S_1$** | Mn(III,IV,IV,III) | **$S_1$** | Mn(III,IV,IV,III) | XFEL |
| 5ws6[45] | $S_3$ | $S_1$ | Mn(III,IV,IV,III) | $S_1$ | Mn(III,IV,IV,III) | XFEL |
| 5zzn[48] | $S_1$ | $S_{-1}$ | Mn(III,IV,III,II) | $S_{-1}$ | Mn(III,IV,III,II) | cryoEM |
| 6dhe[49] | $S_1$ | **$S_1$** | Mn(III,IV,IV,III) | **$S_1$** | Mn(III,IV,IV,III) | XFEL |
| 6dhf[49] | $S_2$ | **$S_2$** | Mn(III,IV,IV,IV) | $S_1$ | Mn(III,IV,IV,III) | XFEL |
| 6dho[49] | $S_3$ | $S_2$ | Mn(IV,IV,IV,III) | $S_1$ | Mn(III,IV,IV,III) | XFEL |
| 6dhp[49] | $S_0$ | $S_3$ | Mn(IV,IV,IV,IV) | $S_2$ | Mn(III,IV,IV,IV) | XFEL |
| 6jlj[50] | $S_1$ | **$S_1$** | Mn(III,IV,IV,III) | **$S_1$** | Mn(III,IV,IV,III) | XFEL |
| 6jlk[50] | $S_2$ | $S_1$ | Mn(III,IV,IV,III) | $S_1$ | Mn(III,IV,IV,III) | XFEL |
| 6jll[50] | $S_3$ | $S_2$ | Mn(IV,IV,IV,III) | $S_2$ | Mn(IV,IV,IV,III) | XFEL |
| 6jlm[50] | $S_1$ | **$S_1$** | Mn(III,IV,IV,III) | **$S_1$** | Mn(III,IV,IV,III) | XFEL |
| 6jln[50] | $S_2$ | $S_0$ | Mn(III,IV,III,III) | $S_1$ | Mn(III,IV,IV,III) | XFEL |
| 6jlo[50] | $S_3$ | $S_0$ | Mn(III,III,IV,III) | $S_1$ | Mn(III,IV,IV,III) | XFEL |
| 6jlp[50] | $S_0$ | $S_1$ | Mn(IV,IV,IV,II) | $S_1$ | Mn(III,IV,IV,III) | XFEL |
| 6w1o[19] | $S_1$ | **$S_1$** | Mn(III,IV,IV,III) | **$S_1$** | Mn(III,IV,IV,III) | XFEL |
| 6w1p[19] | $S_2$ | $S_1$ | Mn(III,IV,IV,III) | $S_1$ | Mn(III,IV,IV,III) | XFEL |
| 6w1v[19] | $S_3$ | **$S_3$** | Mn(IV,IV,IV,IV) | $S_1$ | Mn(III,IV,IV,III) | XFEL |
| 7cji[51] | $S_1$ | **$S_1$** | Mn(III,IV,IV,III) | $S_0$ | Mn(III,IV,IV,II) | XFEL |
| 7cjj[51] | $S_2$ | $S_1$ | Mn(III,IV,IV,III) | $S_1$ | Mn(III,IV,IV,III) | XFEL |
| 7cou[51] | $S_1$ | **$S_1$** | Mn(III,IV,IV,III) | $S_0$ | Mn(III,IV,IV,II) | XFEL |
| 7d1t[30] | $S_1$ | $S_{-4}$ | Mn(III,II,II,II) | $S_{-4}$ | Mn(III,II,II,II) | cryoEM |
| 7d1u[30] | $S_1$ | $S_{-2}$ | Mn(III,III,III,II) | $S_{-2}$ | Mn(III,III,III,II) | cryoEM |
| 7n8o[40] | $S_1$ | $S_{-5}$ | Mn(II,II,II,II) | $S_{-5}$ | Mn(II,II,II,II) | cryoEM |
| 7rcv[40] | $S_1$ | $S_{-5}$ | Mn(II,II,II,II) | $S_{-5}$ | Mn(II,II,II,II) | cryoEM |
| 7rf2[52] | $S_1$ | $S_2$ | Mn(III,IV,IV,IV) | **$S_1$** | Mn(III,IV,IV,III) | XFEL |
| 7rf3[52] | $S_2$ | $S_1$ | Mn(III,IV,IV,III) | **$S_2$** | Mn(III,IV,IV,IV) | XFEL |
| 7rf8[52] | $S_3$ | $S_2$ | Mn(III,IV,IV,IV) | $S_1$ | Mn(III,IV,IV,III) | XFEL |

Table 1. The reported vs. predicted S-states of the X-ray, XFEL and cryoEM structures.

The PDB IDs are sorted alphabetically. The bolded entries are for the structures that the predicted S-state matched the reported

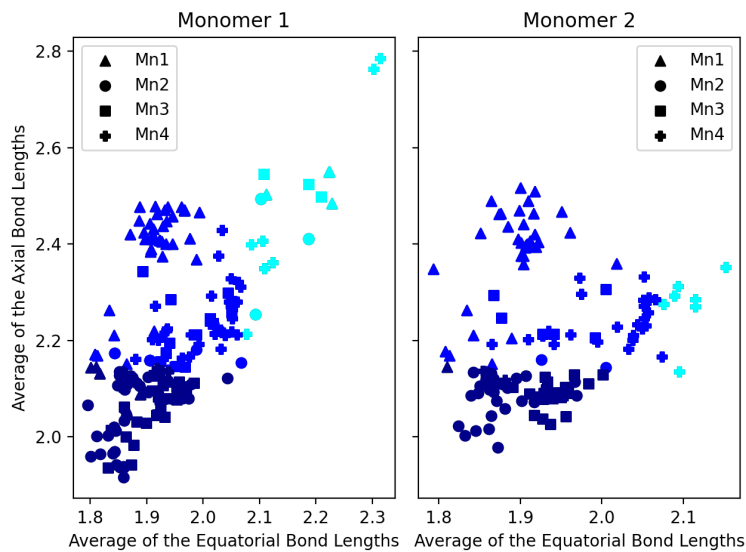

**Fig. 6** The distribution of the oxidation states of the 4 Mn ions in the OEC (Mn(II): cyan, Mn(III): blue, Mn(IV): dark blue) for monomers 1 and 2.

According to our model, Mn2, Mn3 are mostly in the Mn(IV) oxidation state, while Mn1 and Mn4 is mostly in the Mn(III) states (Fig. 6), which agrees with several theoretical studies proposed that Mn2 and Mn3 are oxidized in the $S_1$ state. The transition to $S_2$ takes place by likely oxidizing either Mn4 (as in 6dhf) or Mn1 (as in 6dho). According to theoretical studies supported by EPR measurements, the oxidation of Mn4 in the S2 is responsible for the g=2 EPR signal, while the multi-lines g=4.1 signal may be attributed to the oxidation of Mn1 or the conversion of O4 from a µ-oxo to hydroxo bridge.[53] Furthermore, several studies suggested that the transition between the $S_1$ and the $S_3$ will involve the oxidation of Mn4, which is then reduced and Mn1 is oxidized before both Mn are oxidized in the $S_3$ state. It is interesting to notice that this sequence of the reaction is observed in the structures resolved by *Kern et. al.* (6dhe, 6dhf, 6dho, 6dhp)[49] which are predicted to be in the $S_1$, $S_{2,Mn4(IV)}$, $S_{2,Mn1(IV)}$ and $S_3$ states (Table1).[28, 54, 55] Although, the 6dhp structure is assigned for $S_0$ the existence of the additional water ligand near Mn1 support the prediction by our model.

**Conclusion**

In conclusion, we built two models based on the Decision Tree Classifier (DT) and the Gaussian Naïve Bayes Classifier (GNB) to predict the oxidation state of the Mn ions in the OEC using the available small molecules from the Cambridge Structure Data. The DT model showed better results than the GNB model; it has an accuracy of nearly 100% in the prediction of small molecules and ~ 75% in the case of XFEL-$S_1$ structures. Furthermore, the prediction of the synchrotron and cryoEM $S_1$ structures predicted reduced structures ($S_0$, $S_{-1}$, …, $S_{-5}$), indicating the presence of severe radiation damage, in agreement with several studies that suggested the presence of radiation damage during data collection. The cryoEM structures, in particular, showed significantly high reduced states, up to $S_{-5}$. The observation of radiation damage signs emphasizes the importance of radiation-free data collection to investigate the functional OEC of PSII. In addition, the model predicted that Mn1 and Mn4 are more likely to be oxidized during the transitions $S_1 \rightarrow S_2$ and $S_2 \rightarrow S_3$ states. Moreover, the prediction model shows that Mn4 is the most susceptible Mn ion among the four ions to radiation damage.

**Methods**

Data Collection

We used ConQuest software to search the Cambridge Structural Database for small molecules that contain Mn ions. The first quest resulted in more than 15000 small molecules; however, several filters were used to end up with a reliable data set that resembles some features of the OEC. We started by eliminating the noncrystallographic structures, also any structure with R-factors $\geq 0.075$. Another filter was added to improve the preciseness of the bond length in the structures by including only the error-free structures (at the level of 0.05Å).[22] Furthermore, only Mn compounds with oxygen (O) and nitrogen (N) ligands were included. Finally, overall, we built a database of thousands of octahedral coordination compounds containing 1734, 835, and 107 structures corresponding to the oxidation states Mn(II), Mn(III), and Mn(IV), respectively.

Machine Learning Models

We used sklearn[56] to build prediction models using Gaussian Naïve Bayes and Decision Tree classifiers based on two features: 1) the average distances from Mn to the equatorial and 2) axial ligands. In-house python scripts are written to read the PDB files, extract the octahedral Mn ions, and calculate the distances between the Mn and the ligands using the BioPython[57] package. Then, a CSV file that contains the extracted data from all PDB files is created. Pandas library is used to parse the input data to the machine learning models. The K-mean clustering algorithm in sklearn is used to cluster the Mn ions based on the average bond length of the equatorial and axial ligands into three clusters. Each cluster represents a different oxidation state of the Mn; then, the supervised learning process based on Gaussian Naïve Bayes and Decision Tree classifiers is repeated based on the clustered data.

**Acknowledgments**

We thank Dr. Mohamed Ibrahim, Prof. Dr. Christian Limberg and Dr. Beatrice Cula (Institute of Chemistry, Humboldt Universität zu Berlin) for supporting the CSD data collection. We acknowledges the support from the DOE Grants DESC0001423 (M.R.G. and V.S.B.). We thank Prof. Gary Brudvig and Prof. Marilyn Gunner for the useful discussion.